\documentclass[twocolumn, twocolappendix]{aastex631}
\usepackage{graphicx}

\shorttitle{Unveiling frequency-dependent eclipsing with broadband polar observations}
\shortauthors{Kumari et al.}

\begin{document}
\title{Unveiling frequency-dependent eclipsing in spider millisecond pulsars using broadband polarization observations with the Parkes}
\correspondingauthor{Sangita Kumari}
\email{skumari@ncra.tifr.res.in}

\author[0000-0002-3764-9204]{Sangita Kumari}
\affiliation{National Centre for Radio Astrophysics, Tata Institute of Fundamental Research, S. P. Pune University Campus, Pune 411007, India}
\author[0000-0002-6287-6900]{Bhaswati Bhattacharyya}
\affiliation{National Centre for Radio Astrophysics, Tata Institute of Fundamental Research, S. P. Pune University Campus, Pune 411007, India}
\author[0009-0001-9428-6235]{Rahul Sharan}
\affiliation{National Centre for Radio Astrophysics, Tata Institute of Fundamental Research, S. P. Pune University Campus, Pune 411007, India}
\author[0000-0002-7122-4963]{Simon Johnston}
\affiliation{CSIRO Astronomy and Space Science, P.O. Box 76, Epping, NSW 1710, Australia}
\author[0000-0003-2122-4540]{Patrick Weltevrede}
\affiliation{Jodrell Bank Centre for Astrophysics, Department of Physics and Astronomy, The University of Manchester, UK}
\author[0000-0001-9242-7041]{Benjamin Stappers}
\affiliation{Jodrell Bank Centre for Astrophysics, Department of Physics and Astronomy, The University of Manchester, UK}
\author[0000-0001-8801-9635]{Devojyoti Kansabanik}
\affiliation{Cooperative Programs for the Advancement of Earth System Science, University Corporation for Atmospheric Research, Boulder, CO, USA}
\affiliation{NASA Jack Eddy fellow at the Johns Hopkins University Applied Physics Laboratory, 11001 Johns Hopkins Road, Laurel, MD 20723, USA}
\affiliation{National Centre for Radio Astrophysics, Tata Institute of Fundamental Research, S. P. Pune University Campus, Pune 411007, India}
\author[0000-0002-2892-8025]{Jayanta Roy}
\affiliation{National Centre for Radio Astrophysics, Tata Institute of Fundamental Research, S. P. Pune University Campus, Pune 411007, India}
\author[0009-0002-3211-4865]{Ankita Ghosh}
\affiliation{National Centre for Radio Astrophysics, Tata Institute of Fundamental Research, S. P. Pune University Campus, Pune 411007, India}

\begin{abstract}
This study presents an orbital phase-dependent analysis of three black widow spider millisecond pulsars (BW MSPs), aiming to investigate the magnetic field within the eclipse environment. The ultra-wide-bandwidth low-frequency receiver (UWL) of the Parkes ‘Murriyang’ radio telescope is utilized for full polarization observations covering frequencies from 704$-$4032 MHz.
Depolarization of pulsed emission is observed during the eclipse phase of three BW MSPs namely, J0024$-$7204J, J1431$-$4715 and PSR J1959+2048, consistent with previous studies of other BW MSPs. 
We estimated orbital phase dependent RM values for these MSPs. The wide bandwidth observations also provided the constraints on eclipse cutoff frequency for these BW MSPs. For PSR J0024$-$7204J, we report temporal variation of the eclipse cutoff frequency coupled with changes in the electron column density within the eclipse medium across six observed eclipses. Moreover, the eclipse cutoff frequency for PSR J1431$-$4715 is determined to be 1251 $\pm$ 80 MHz, leading to the conclusion that synchrotron absorption is the primary mechanism responsible for the eclipsing. Additionally, for PSR J1959$+$2048, the estimated cutoff frequency exceeded 1400 MHz, consistent with previous studies. With this investigation, we have doubled the sample size of BW MSPs with orbital phase-resolved studies allowing a better probe to the eclipse environment. 

\end{abstract}

\keywords{pulsars: general; binaries: eclipsing, pulsars: individual}

\section{Introduction}
\label{sec:intro}


Eclipsing spider millisecond pulsars (MSPs) are compact binary pulsar systems in which the energetic pulsar wind ablates the companion and this ablated material is assumed to be the cause of observed eclipses \citep[e.g. $\sim 10\%$ of the binary orbit is eclipsed for PSR J1959$+$2048,][]{Fruchter1988a}. Since the radius of the Roche lobe of the companion \citep{Roche_eggleton} is smaller than the observed eclipse duration, the material causing the eclipse is not gravitationally bound to the companion star \citep[e.g.][]{Roche_lobe_eclipse,RochelobeJ2055}. Based on the mass of the companion ($\mathrm{M_{c}}$), spider MSPs are divided into two categories: black widow \citep[BW, $\mathrm{M_{c}}< 0.05 \mathrm{M}_{\odot}$,][]{roberts2012surrounded} and redback \citep[RB, $0.1 \mathrm{M}_{\odot} < \mathrm{M_{c}}< 0.9 \mathrm{M}_\odot$,][]{roberts2012surrounded}. Owing to the frequency-dependent nature of the eclipses, the pulsed signal from the pulsar disappears within the eclipse region below a certain frequency called the eclipse cutoff frequency \citep[$\nu_{c}$,][]{kansabanik2021unraveling}. 

The observed properties of pulsed radio emission are affected by intervening magnetized plasma, either by the interstellar medium (ISM) or by the eclipsing medium. Due to the electron column density, pulsed emission at lower frequency is delayed compared to higher frequency emission. Under the assumption $\mathrm{f_{p}=8.98 \, kHz \times \sqrt{\frac{n_{e}}{cm^{-3}}}<< f}$ and $\mathrm{f_{c}=2.80 \, MHz \times\frac{B}{1G}<<f}$, where $\mathrm{f_{p}}$ is the plasma frequency, $\mathrm{f_{c}}$ is the cyclotron frequency and $\mathrm{f}$ is the observing frequency, this time delay (t) is proportional to a quantity known as dispersion measure \citep[DM,][]{LorimerKramer},
\begin{equation}
\label{eqn:DMdeterimation}
    \mathrm{DM (pc~cm^{-3}) = 2.4 \times 10^{-10}t(\mu s)f^{2}(MHz)} 
\end{equation}
If the pulsed emission is linearly polarized, which is observed for most of the normal pulsars and MSPs, the polarization angle rotates as it passes through the magnetized plasma. This rotation is called Faraday Rotation and is characterized by a frequency independent quantity called, Rotation Measure \citep[RM,][]{LorimerKramer},
\begin{equation}
\mathrm{RM (rad~m^{-2}) = 0.810 \int_{0}^{D} n_{e}}\vec{B}.\vec{dl}
\end{equation}
where, $n_{e}$ is the ISM electron density in units of $cm^{-3}$, $\vec{B}$ is the magnetic field in microgauss and $\vec{dl}$ is a minute directional component aligned with the line of sight directed towards us in parsecs.  Assuming the uniform electron density and magnetic field along the line of sight, the average value of the magnetic field along the direction of wave propagation is described by the following relation,
\begin{equation}
\label{magnetic_determination}
\mathrm{<B_{||}(\mu G)>~ = 1.232 \left(\frac{RM}{rad~m^{-2}}\right)\left(\frac{DM}{cm^{-3}pc}\right)^{-1}}   
\end{equation}

Polarization studies in BW MSPs exhibiting eclipses are crucial for determining the magnetic field in the eclipse medium, which plays a significant role in understanding the eclipse mechanism \citep[e.g.][]{PolzinJ2051,J1748depol}. An in-depth exploration of potential eclipse mechanisms is provided by \cite{Thompson1992}, offering explanations for the frequency-dependent eclipses in spider MSP systems. \cite{Thompson1992} highlighted that different systems may have different eclipse mechanisms.  \cite{PolzinJ1810}, \cite{Kudale2020} and \cite{kansabanik2021unraveling} found cyclotron-synchrotron absorption to be the primary mechanism for eclipses in PSR J1810$+$1744, PSR J1227$-$4853 and PSR J1544$+$4937 respectively. Scattering and cyclotron absorption are considered as the main mechanisms for PSR J2051$-$0827 \citep{PolzinJ2051}. Additionally, stimulated Raman scattering has been proposed as the most plausible eclipse mechanism for PSR B1744$-$24A \citep{Thompson1992}. By estimating the magnetic field at the superior conjunction ($\sim$ orbital phase 0.25), one can gain valuable insights into whether cyclotron or synchrotron absorption is responsible for the observed eclipses in spider systems.

Conducting RM determinations at the eclipse centre would also aid in better understanding the temporal evolution of the magnetic field in the eclipse medium. This understanding would be valuable for studying the causes of variation in the eclipse cutoff frequency in spider systems. Temporal variations in the eclipse cutoff frequency have been observed for PSR J1544+4937 by \cite{kumari2024}. The study by \cite{kumari2024} proposed that the temporal changes in the magnetic field within the eclipse environment could contribute to the variation in the eclipse cutoff frequency, along with variations in parameters such as electron density, the spectral index of electron energy distribution, absorption length, and the angle of the magnetic field along our line of sight. Therefore, RM measurements in the eclipse environment would also help to differentiate the effects of the magnetic field on the cutoff frequency from other factors. This differentiation would aid in investigating the possible reasons for changes in the eclipse cutoff frequency for BW MSPs.

Polarization studies have been conducted only in a handful of eclipsing BW MSP systems. Depolarization is consistently observed around superior conjunction in these systems, such as PSR J1748$-$2446A \citep{J1748_nature}, J2051$-$0827 \citep{PolzinJ2051,J2051depol}, and J2256$-$1024 \citep{J2256depol}. Additionally, changes in the RM value at the eclipse boundary are noted for PSR J2256$-$1024 \citep{J2256depol} and J2051$-$0827 \citep{J2051depol}.
Thus, although the importance of orbital phase-dependent polarization studies is evident, this domain remains largely unexplored, having been conducted for only $\sim$ 4$\%$ of BW MSPs. Motivated by this we have conducted the orbital phase dependent study for 3 BW MSPs using the Parkes ‘Murriyang’ radio telescope ultra-wide-bandwidth low-frequency receiver \citep[UWL, covering the frequency range from 704 MHz$-$4032 MHz,][]{UWL_receiver} data. The UWL at the Parkes offers a wide bandwidth for identifying the cutoff frequency and studying the corresponding temporal variations in the spider system. Simultaneous polarization observations also provide a platform to study the orbital phase dependent measurement of the magnetic field using RM analysis in these systems.

The structure of this paper is as follows, Section \ref{sec:obs} presents the details of the observations and data analysis. The results are presented in Section \ref{results}, followed by a discussion in Section \ref{discussion}. Section \ref{summary} provides a summary of the work.

\begin{table*}[!htb]
\begin{center}
\caption{Summary of the observations$^{\dagger}$}
\label{tab:Table1}
\vspace{0.3cm}
\label{discovery}
\begin{tabular}{|l|l|l|l|l|l|l|l|l|l}
\hline
Pulsar name     & P$^{a}$ & DM$^{b}$ & $P_{b}^{c}$& Proposal code & Date & $T_{int}^{d}$ & OP$^{e}$ & $T_{sub}^{f}$\\
     & (ms)& ($pc~cm^{-3}$)& (hrs) & & & (hrs) &  & (s)\\
 
\hline
J0024$-$7204J   &2.101 &24.59& 2.8 &P1078$^{g}$ &16$-$08$-$2020 & 5.0 &  0.22$-$1.98$^{*}$ & 30\\
&&& & &27$-$01$-$2021 & 3.3 &    0.49$-$1.62$^{*}$  & 30  \\
 
&&&  & P1006 & 29$-$05$-$2019  & 3.6 &     0.39$-$1.68$^{*}$  & 20 \\
  
&&&   & & 08$-$07$-$2019 &  3.1 &    0.20$-$1.27$^{*}$  & 20 \\
&&&    & & 14$-$07$-$2019 & 2.8 &   0.97$-$1.94$^{*}$  & 20   \\
\hline
J1431$-$4715  &2.012&59.35&10.7&P789& 23$-$11$-$2018  & 3.6 &  0.62$-$0.94  & 20   \\
 &&& && 30$-$11$-$2018 &  0.4 &  0.40$-$0.45   &   20  \\
  &&&  && 08$-$12$-$2018 & 0.7 &   0.96$-$0.00      & 30  \\
   &&&   && 13$-$02$-$2019 & 0.4 &  0.07$-$0.26  & 30         \\
   &&&     && 27$-$04$-$2019 & 1.2 &   0.81$-$0.93   & 10      \\
   &&&       && 07$-$06$-$2019 & 1.6 & 0.28$-$0.42    &  10     \\
\hline

J1959$+$2048 &1.607&29.11& 9.1& P1078$^{g}$ & 06$-$10$-$2020  &2.4 &0.07$-$0.32 & 30  \\
&&&  & & 16$-$10$-$2020    & 2.8 & 0.12$-$0.43  & 30 \\
&&&  && 10$-$02$-$2021  & 3.1  & 0.17$-$0.51 & 30 \\
   
&&&     & & 27$-$01$-$2021  & 2.8   & 0.03$-$0.35 & 30 \\
\hline

\end{tabular}
\end{center}
$^{a}$: Spin period of the pulsar; $^{b}$: Dispersion measure; $^{c}$: Orbital period of the pulsar; $^{d}$: Total integration time; $^{e}$: Orbital phase covered; $^{f}$: Subintegration length; $^{g}$: Data for P1078 is acquired by some of the authors of this paper; $^{*}$: 1 in front of the orbital phase represents the orbital phase of next consecutive orbit; $^{\dagger}$: Publicly available Parkes UWL data, at CSIRO data archive

\vspace{1cm}
\end{table*}

\section{Observations and data analysis}
\label{sec:obs}

We have selected three BW MSPs namely, J0024$-$7204J, J1431$-$4715 and J1959$+$2048 for which no orbital phase-dependent polarization study has been reported previously. We used data from the Parkes radio telescope, made publicly available through the CSIRO data archive\footnote{https://data.csiro.au}. A summary of these observations is provided in Table 1. The data were taken using the UWL which operates between 704 and 4032 MHz. The data were coherently de-dispersed, had a frequency resolution of 1 MHz, and were folded into 1024 phase bins at the pulsar's topocentric period and integrated over a sub-integration time as listed in Table 1. Before each observation of the pulsar, a short observation of a pulsed calibration signal injected into the low-noise amplifiers was made. 

We used PSRPYPE\footnote{https://github.com/vivekvenkris/psrpype} for cleaning the data. Flux calibration is conducted using observations of the source PKS~1934$-$638, which converts the digitiser units to mJy. We used $\it{PSRCHIVE}$ task “$\it{pac}$” to apply the flux and polarization calibrator solutions to our dataset. The method outlined in \cite{pcm_file} is used to account for the instrumental leakage terms. Thereafter the remaining RFI is mitigated in time and frequency domain utilizing the $\it{PSRCHIVE}$ task “$\it{pazi}$”. To increase the detection significance we averaged across frequency and phase bins by factors of 4 and 2, respectively, using the $\it{PSRCHIVE}$ task “$\it{pam}$”. This resulted in a calibrated profile per sub-integration comprising 4 Stokes parameters, 832 frequency channels, and 512 phase bins for each pulsar. This configuration was used for further analysis. All consecutive observations within a specific epoch were combined using the $\it{PSRCHIVE}$ task “$\it{psradd}$”, generating a single output file. A time drift observed in the combined file was rectified by updating the parameter file. This involved fitting the time of arrivals (TOAs) calculated using the $\it{PSRCHIVE}$ task “$\it{pat}$”, solely for spin frequency (F0), epoch of ascending node (T0), and projected semi-major axis (A1) using $\it{tempo2}$ \citep{hobbs2006tempo2}. 

To find the eclipse cutoff frequency for all the MSPs in our sample, we first divided the observing bandwidth into 160 MHz chunks and searched for pulsed signals within the eclipse region for each pulsar. The eclipse region is taken to be different for different pulsars, and the details of each are given in Table \ref{table2}. Both OFF and ON phase bins follow Gaussian distribution with the same standard deviation but with different mean, in the baseline subtracted data cube (time, frequency, phase bin). To determine the ON and OFF phase bins, a quantity H = (sum of N number of samples)/(standard deviation $\times \sqrt{tf}$) (Sharan et al. 2024, in preparation) is calculated for each phase bin, where N = $tf$ is the number of samples, t is the number of sub-integrations in time during the eclipse phase, and f corresponds to the number of sub-bands in frequency for a 160 MHz chunk. The phase bins with a value of H greater than 4 were labelled as ON otherwise OFF. Subsequently, based on ON bins the eclipse cutoff frequency is estimated. This method to determine the cutoff frequency has also been used by \cite{kumari2024}.

The RM is determined using the $\it{PSRCHIVE}$ task “$\it{rmfit}$”. The RM estimation for PSR J0024$-$7204J, J1431$-$4715, and J1959$+$2048 is performed over intervals of 8 minutes ($\sim$ 0.05 orbital phase), 19 minutes ($\sim$ 0.03 orbital phase), and 16 minutes ($\sim$ 0.03 orbital phase) in the frequency ranges of 704$-$1050 MHz, 704$-$3500 MHz, and 704$-$1900 MHz, respectively, with a frequency resolution of 4 MHz. The time interval chosen for RM estimation is selected such that within this interval, there is a sufficient fraction of linear polarization in the chosen frequency band to provide a reliable estimate of RM. Previous studies by \cite{Parkes_UWL_RM_J0024}, \cite{Meerkat_J1431_RM} and \cite{J1959_fast_pol_21} have estimated RMs of +20 rad m$^{-2}$, +13.3 rad m$^{-2}$ and $-$70 rad m$^{-2}$ for PSRs J0024$-$7204J, J1431$-$4715 and J1959$+$2048 respectively. Our determination of the RM is not hindered by frequency resolution, as the channel resolution of 4 MHz at 877 MHz, 2.1 GHz and 1.5 GHz would enable RM measurement up to $f^{3}/c^{2}\Delta f$ = 1873 rad m$^{-2}$, 25725 rad m$^{-2}$ and 9375 rad m$^{-2}$ for PSR J0024$-$7204J, J1431$-$4715 and J1959$+$2048, respectively.
The RM correction is done utilizing the $\it{PSRCHIVE}$ task “$\it{pam}$”.

We computed the orbital phase dependent mean flux density by calculating the area under the integrated pulse for specific orbital phase intervals and frequency ranges, divided by the number of phase bins \citep{LorimerKramer}.
The flux density estimation for PSR J0024$-$7204J, J1431$-$4715, and J1959$+$2048 is performed over 60 seconds ($\sim$ 0.003 orbital phase interval), 19 minutes ($\sim$ 0.03 orbital phase interval), and 16 minutes ($\sim$ 0.03 orbital phase interval) time intervals in the 704$-$1107 MHz, 1251$-$1600 MHz, and 704$-$1107 MHz frequency range, respectively.

The excess DM as the function of orbital phase ($DM_{excess}$) is determined from the excess time delay in the pulse arrival time (obtained using $\it{tempo2}$) from Equation \ref{eqn:DMdeterimation}. The corresponding electron column density ($N_{e}$) is then determined using the relation,
\begin{equation}
\label{N_e_determination}
    \mathrm{N_{e}(cm^{-2}) = 3 \times 10^{18} \times DM_{excess}(pc~cm^{-3})}
\end{equation}
The estimated values of the cutoff frequency and peak $N_{e}$ are mentioned in Table \ref{table2}. 


\begin{figure}
     \centering
     \begin{minipage}[b]{0.44\textwidth}
         \centering
         \includegraphics[width=\textwidth]{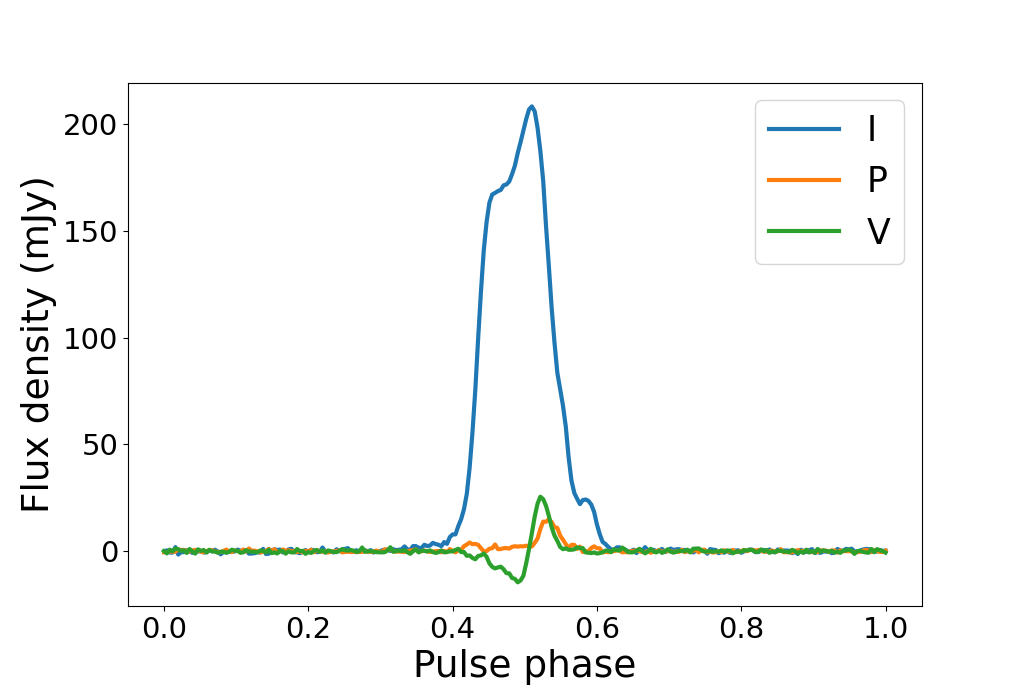}
         \textbf{(a)} J0024--7204J
         \label{J0024_pol}
     \end{minipage}
     \hfill
     \begin{minipage}[b]{0.44\textwidth}
         \centering
         \includegraphics[width=\textwidth]{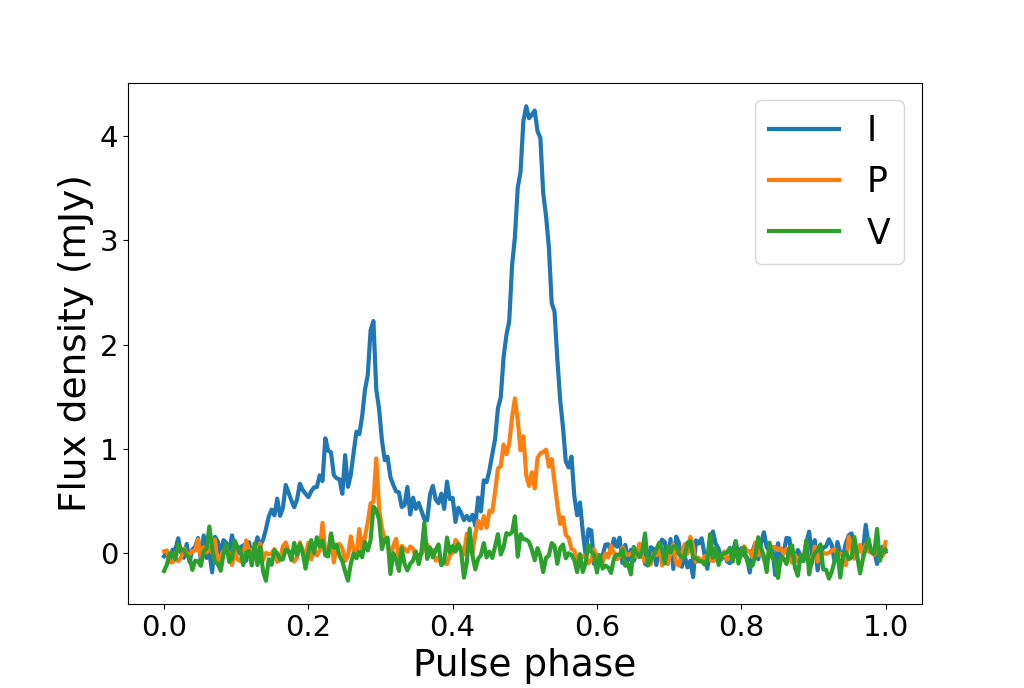}
        \textbf{(b)} J1431--4715
         \label{J1431_pol}
     \end{minipage}
     \hfill
     \begin{minipage}[b]{0.44\textwidth}
         \centering
         \includegraphics[width=\textwidth]{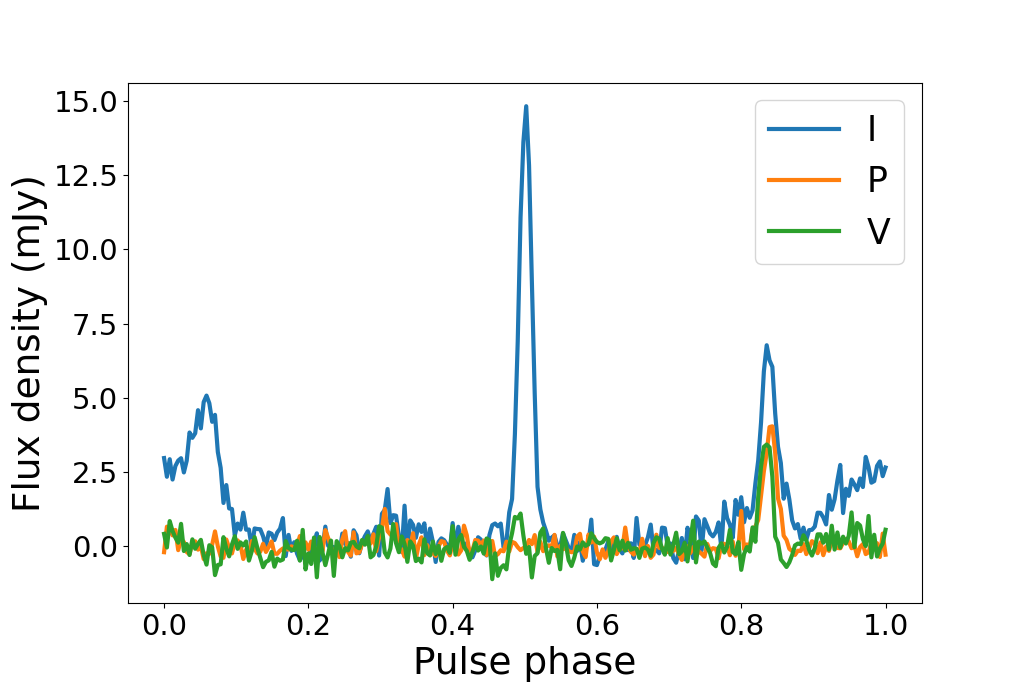}
         \textbf{(c)} J1959+2048
         \label{J1959_pol}
     \end{minipage}
    \caption{The calibrated, RM corrected integrated profiles for PSR J0024$-$7204J (8th July 2019, 704$-$1050 MHz), PSR J1431$-$4715 (23rd November 2018, 704$-$3500 MHz), and PSR J1959$+$2048 (27th January 2021, 704$-$1900 MHz) are depicted. The blue, orange, and green curves represent the total intensity (I), linear polarized intensity (P), and circular polarized intensity (V), respectively.}
        \label{polarisation profile}
\end{figure}

\section{Results}
\label{results}
 
The calibrated and RM corrected integrated profiles for the three MSPs in our sample during the non-eclipse phase are illustrated in Figure \ref{polarisation profile}. For PSR J0024$-$7204J, polarization was observed only on 8th July 2019, with no linear polarization detected in other epochs. Figure \ref{polarisation profile}a displays the integrated profile for this MSP for 8th July 2019 for 704$-$1050 MHz, where we observed 6$\%$ mean circular polarization and 9$\%$ mean linear polarization. The polarization pulse profile for this epoch is consistent with what is reported by \cite{Parkes_UWL_RM_J0024}, as the same dataset is utilized in our analysis. PSR J1431$-$4715 exhibits consistent polarization across all epochs during the non-eclipse phase. A previous study by \cite{Meerkat_J1431_RM} with MeerKAT telescope reported 33$\%$ linear polarization for PSR J1431$-$4715 in the 856$-$1712 MHz frequency range. Figure \ref{polarisation profile}b depicts the pulse profile for PSR J1431$-$4715 on 23 November 2018 for 704$-$3500 MHz, indicating 29$\%$ linear polarization fraction for the main component for this epoch. PSR J1959$+$2048 displays multiple components in its profile, as evident in Figure \ref{polarisation profile}c. No linear polarization was observed for the main component throughout the band. However, 40$\%$ mean linear polarization was observed for the other component (at pulse phase $\sim 0.82$, referred as component 2 hereafter) during the observation on 27 January 2021 in 704$-$1900 MHz frequency range in non-eclipse phase. For this same component 26$\%$ mean circular polarization was also observed.
A previous study by \cite{J1959_fast_pol_21} using the FAST \citep{FAST_telescope} reported 25$\%$ linear polarization for PSR J1959+2048 in the frequency range of 1–1.5 GHz. From Figure A7 of their study, we note a significant linear polarization for component 2, but a low linear polarization fraction for the main component.

\subsection{Cutoff frequency estimates}
\label{cutoff frequency estimates} 
 
A precise estimation of the eclipse cutoff frequency would help to probe the eclipse mechanism responsible for frequency-dependent eclipsing \citep{kansabanik2021unraveling}. We determined the $\nu_{c}$ for the three BW MSPs in our sample, using a 160 MHz frequency chunk as detailed in Section \ref{sec:obs}. For PSR J0024$-$7204J, we obtained the constraints on the cutoff frequency for five epochs listed in Table \ref{table2}. The cutoff frequency for two epochs namely 16 July 2020 and 08 June 2019 is found to be below 734 MHz (the lowest edge of the observing frequency band). However, for another two epochs namely, 29 June 2019 and 27 January 2021 the cutoff frequency could potentially fall in the lower half of the band. The precise determination of the cutoff frequency for the later two epochs is hindered by sensitivity, as the pulsar remains undetected both during the eclipse and non-eclipse phases within the specified frequency range. We got a lower limit of $\nu_{c}$ of 1581 MHz for the 14th July 2019 eclipse, as the pulsar is detected up to 1581 MHz in the non-eclipse phase, but not detected afterwards in the eclipse as well as non-eclipse phase.
The previous study by \cite {J0024_previous_study} reported a complete eclipse at 440 MHz, a variable eclipse at 660 MHz, and no detectable eclipse at 1400 MHz for PSR J0024$-$7204J, suggesting a temporal change in the eclipse cutoff frequency for this system.

For PSR J1431$-$4715, data from eight observing epochs were analyzed, and the eclipse phase is covered only on 13 February 2019 and 07 June 2019, as can be seen from Table \ref{tab:Table1}. We determined the eclipse cutoff frequency for this epoch to be 1251$\pm$80 MHz. 
In the previous study by \cite{J1431_previous_study}, a signature of eclipsing is observed in the timing residuals at 700 MHz and 1400 MHz, but the pulsed emission never disappeared in the eclipse region. Therefore according to \cite{J1431_previous_study} the eclipse cutoff frequency, is expected to remain below 700 MHz for PSR J1431$-$4715. Additionally, our estimate of a significantly higher value suggests a temporal change in $\nu_{c}$ for this MSP. Similarly, for PSR J0024$-$7204J different constraints on $\nu_{c}$ across different epochs, indicate the temporal evolution of the eclipse cutoff frequency. However, the determination of the cutoff frequency could be affected by radio frequency interference, frequency-dependent sensitivity, and the method used for estimation. Temporal changes in the cutoff frequency have been reported for PSR J1544$+$4937 by \cite{kumari2024}, depicting the dynamical evolution of the eclipse environment.

For PSR J1959$+$2048, we analyzed four epochs of observations and found almost identical constraints on the cutoff frequency for each epoch. Details are provided in Table \ref{table2}. On 6 October 2020, 16 October 2020 and 27 January  2021, the cutoff frequency is above 1424 MHz. In these three epochs the signal is detected in the non eclipse phase below 1424 MHz. However above 1424 MHz the signal is not detected in both eclipse and non-eclipse phase, which could be due to the pulsar spectra \citep{high_frequency_turnover,high_frequency_turnover_b}. Eclipse up to 1.4 GHz is seen for this pulsar in the previous studies \citep{J1959_previous_study}.


\begin{table*}[hbt!]
\begin{center}
\caption{Table listing the cutoff frequency values for multiple observing epochs along with the corresponding electron column density in the eclipse medium.}
\label{tab:Table3}
\vspace{0.3cm}
\label{table2}
\begin{tabular}{|l|l|l|l|l|l|}
\hline
Pulsar name   & Date & Cutoff frequency($\nu_{c}$)& Eclipse orbital phase$^{a}$ & $N_{e}^{b}$ & Frequency$^{c}$\\
   &  & (MHz)&  & $(cm^{-2})$ & (MHz)\\
\hline
J0024$-$7204J   & 16$-$08$-$2020 & $<$ 734 & 0.22$-$0.26& $3\times 10^{16}$ & 734$-$1107 \\
 & 27$-$01$-$2021 & $<$ 1911 & 0.22$-$0.26& $4\times 10^{16}$& 1911$-$2221\\
  & 29$-$05$-$2019 & $<$ 1232 & 0.22$-$0.26&$5\times 10^{16}$& 1232$-$1655\\
   & 08$-$07$-$2019$^{1}$ & $<$ 734 & 0.18$-$0.21& $1\times 10^{16}$ & 734$-$1107\\
   & 08$-$07$-$2019$^{2}$ & $<$ 734 & 0.18$-$0.21 &$2\times 10^{16}$& 734$-$1107\\
    & 14$-$07$-$2019 & $>$ 1581 &0.22$-$0.26 &$4\times 10^{16*}$& 1251$-$1581\\
\hline
J1431$-$4715 &13$-$02$-$2019 & 1251 $\pm$ 80& 0.22$-$0.25&$9\times 10^{17*}$&1251$-$1600\\
     &07$-$06$-$2019 & 1251$\pm$ 80 & 0.28$-$0.30 &$9\times 10^{17*}$&1251$-$1600\\
\hline
J1959$+$2048 & 06$-$10$-$2020 & $>$ 1424 &0.23$-$0.26 &$4 \times 10^{16*}$&734$-$1107\\
 & 10$-$02$-$2021 & $>$ 1581  & 0.23$-$0.26&$3 \times 10^{17*}$&734$-$1107\\
  & 16$-$11$-$2020 & $>$ 1424 & 0.23$-$0.26&$5 \times 10^{16*}$&734$-$1107\\
   & 27$-$01$-$2021 & $>$ 1424 & 0.23$-$0.26&$1 \times 10^{17*}$&734$-$1107\\

\hline

\end{tabular}
\end{center}
$^{a}$: The orbital phase range averaged to determine the eclipse cutoff frequency.\\
$^{b}$: The maximum electron column density in the eclipse medium.\\
$^{c}$: Frequency chunk that has been considered to find the maximum value of $N_{e}$ in the eclipse medium.\\
$^{*}$: The epochs where only the limit on the $N_{e}$ is known.\\
$^{1}$: 1st orbit \\
$^{2}$: 2nd orbit \\

\vspace{1cm}
\end{table*}

\subsection{Variation of total intensity with orbital phase}
 
We examined the variation in observed intensity as a function of orbital phase for all three MSPs, to investigate the effect of the eclipse medium on the observed flux density. For PSR J0024$-$7204J, only on 16 October 2020, we observed a slight reduction in the flux density (not completely zero) around the superior conjunction ($\sim$ orbital phase 0.25) using only the lower part ($\sim$ 700--1107 MHz) of the UWL band. In other epochs for this pulsar, we did not observe a decrease in the total intensity near orbital phase 0.25.

\begin{figure*}[!htbp]
\begin{center}
\includegraphics[width=0.5\textwidth]{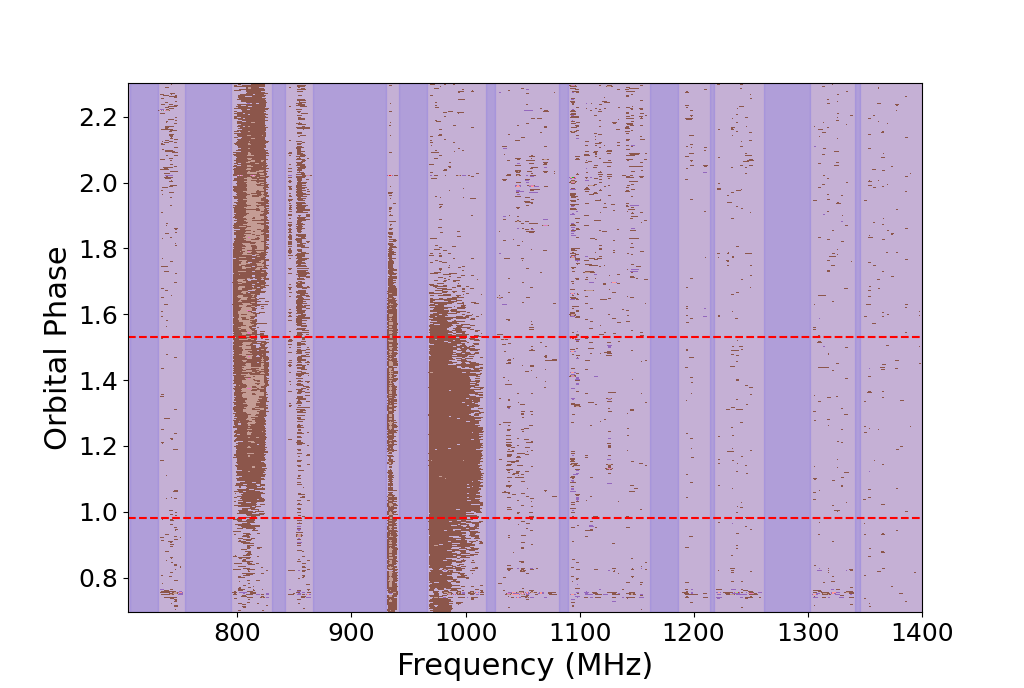}
\caption{The dynamic spectra for PSR J0024$-$7204J on 8 July 2019. The brown colour shows the detection. The MSP is detected only in the lower part of the UWL band (approximately 780-1000 MHz). The flagged frequencies due to RFI are shown by the shaded violet region. The eclipse region, delineated by red dotted lines, is marked based on the observed depolarization width. }
\label{fig:Dynamic_spectra}
\end{center}
\end{figure*}

On July 8 2019, two consecutive eclipses are covered for this pulsar. For the first eclipse, we observed an increase in flux density in the eclipse region, despite the presence of extra material (indicated by $DM_{excess}$ and $N_{e}$) as shown in Figure \ref{fig:J0024_plot}(c). However, for the second eclipse, we did not observe any increase in flux density in the eclipse region.  The flux density in the eclipse phase for the first eclipse is around a factor of 4 higher compared to the non-eclipse phase flux density (evident from Figure \ref{fig:J0024_plot}b). 
The observed flux magnification is chromatic and is seen only for the lower part of the UWL band (between 780--1000 MHz). Figure \ref{fig:Dynamic_spectra}, depicts the dynamic spectra for 8 July 2019. From the dynamic spectra, we inferred that the signal is dominant at an earlier time in the higher frequency chunk (around 980 MHz), while at a later time, the signal is dominant in the lower frequency chunk (around $\sim$800 MHz).
To verify whether this magnification is due to scintillation from the interstellar medium (ISM), we determined the scattering timescale by fitting the integrated profile around the eclipse phase (orbital phase $\sim$ 0.15-0.45) with a function that is a convolution of an exponential and the sum of three Gaussian functions. The obtained scattering timescale ($t_{\text{scat}}$) is 70 $\mu$s for PSR J0024$-$7204J. Consequently, the de-correlation bandwidth is $1/(2 \times 3.14 \times t_{\text{scat}}) = ~2 ~\mathrm{kHz}$, which is less than the frequency resolution (1 MHz) of our data set. Therefore, scintillation from the ISM could not be responsible for this magnification.

The increase in flux density during the eclipse phase could be the first detection of plasma lensing in PSR J0024$-$7204J. This phenomenon may explain why linear polarization is observed specifically during this epoch. Previous studies have reported plasma lensing in  BW MSPs. For example, \cite{J2051_plasma_lensing}, \cite{J1959_plasma_lensing}, and \cite{B1744_plasma_lensing} observed plasma lensing in PSR J2051$-$0827, J1959+2048, and PSR B1744$-$24A, respectively. \cite{J2051_plasma_lensing} observed two consecutive eclipses in their observation, during which flux brightening is observed only for the first eclipse, similar to what we have observed for PSR J0024$-$7204J. 

For PSR J1431$-$4715, we carried out a similar analysis of intensity variation with orbital phase and observed a decrease in intensity in the eclipse region as evident from Figure \ref{fig:J1431_plot}c.  
Similarly, for PSR J1959$+$2048, we observed a reduction in flux density in the 704$-$1107 MHz frequency range for all epochs of observations, around the eclipse. Figure \ref{fig:J1959_plot}b presents the flux density variation with orbital phase for the observation on 27 January 2021, specifically for the component 2.

\subsection{DM variation in the eclipse medium}

\begin{figure*}[!htbp]
\begin{center}
\includegraphics[height=0.45\textheight, width=1.0\textwidth, angle=0]{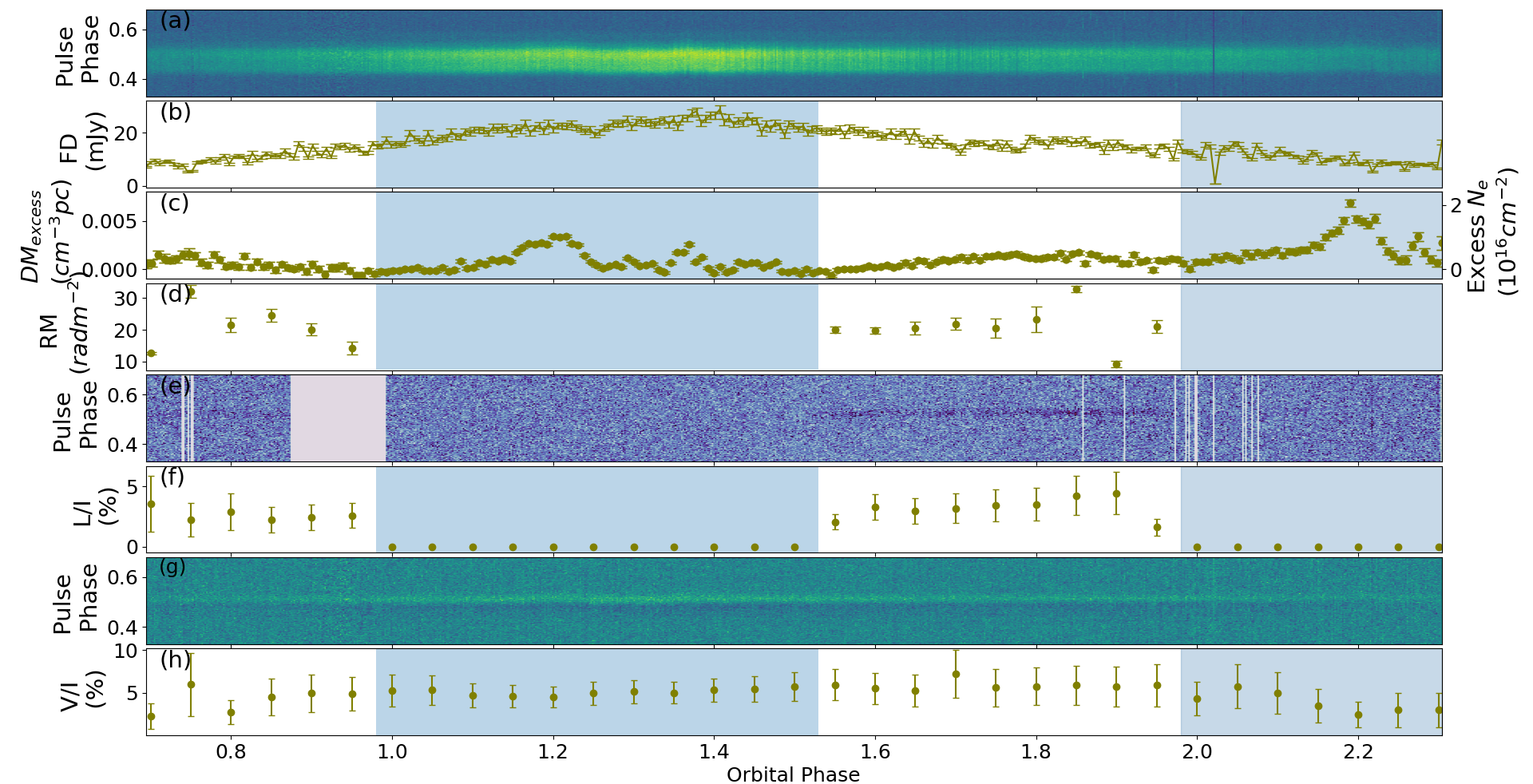}
\caption{Eight panels, from top to bottom depict variation of (a) the total intensity, (b) flux density of total intensity (FD), (c)  excess DM ($DM_{excess}$) along the line of sight, (d) RM, (e) linear polarization intensity, (f) linear polarization percentage, (g) circular polarization intensity, and (h) circular polarization percentage, with orbital phase for PSR J0024$-$7204J on 8 July 2019. The above quantities have been calculated for the 704$-$1107 MHz frequency range. The blue shaded eclipse regions are based on the depolarization width.}
\label{fig:J0024_plot}
\end{center}
\end{figure*}

To map the orbital phase dependent $N_{e}$, we examined the variation of DM, in the lowest possible 350 MHz chunk in frequency which is mentioned for individual sources in the table \ref{table2} (to ensure reasonable detectability, as the signature of eclipsing is prominent at lower frequencies). We observed an increase in the DM value during the eclipse phase (around 0.2 orbital phase) only on 8th July 2019, with a small increase also noted on 16th August 2020. The $DM_{excess}$ as a function of the orbital phase for 8th July 2019 is illustrated in Figure \ref{fig:J0024_plot}c for PSR J0024$-$7204J. In other epochs, no such orbital phase-dependent DM variation is detected. Moreover, in the non-eclipse phase also we observed an increase in DM value for this MSP, indicating possible clumping of material. 

\begin{figure*}[!htbp]
\begin{center}
\includegraphics[width=0.99\textwidth,angle=0]{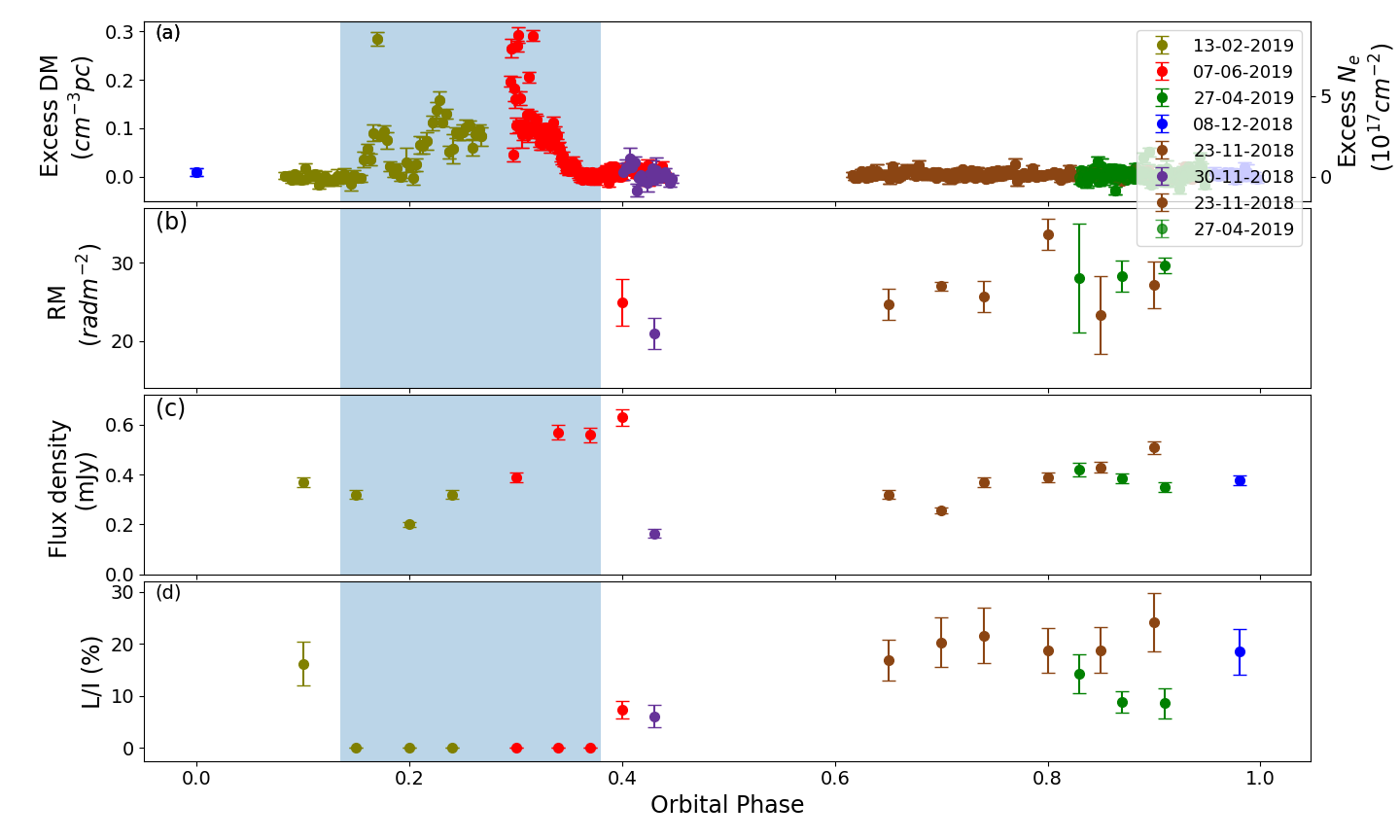}
\caption{Four panels, from top to bottom depict, (a) excess DM ($DM_{excess}$) along the line of sight, (b) RM, (c) flux density of total intensity, (d) linear polarization percentage (of the main component) for PSR J1431$-$4715. Different colours represent different epochs of observations. The $DM_{excess}$, flux density and linear polarization percentage variation are determined using the 1251$-$1600 MHz frequency range, given that the cutoff frequency for this pulsar is 1251 MHz. The blue-shaded eclipse region is based on the depolarization width.}
\label{fig:J1431_plot}
\end{center}
\end{figure*}

For PSR J1431$-$4715, the orbital phase-dependent DM variation is depicted in Figure \ref{fig:J1431_plot}a, over the 1251-1600 MHz frequency range (around the cutoff frequency of the pulsar). An increase in DM around orbital phase 0.25 suggests the presence of additional material. The lower limit of the maximum $DM_\mathrm{{excess}}$ during the eclipse phase is approximately $0.3 ~pc~ cm^{-3}$, and the corresponding excess $N_{e}$ is around $9 \times 10^{17} cm^{-2}$ according to Equation \ref{N_e_determination}. These values represent the lower limit since the observations did not cover the eclipse phase between 0.26 to 0.28, where the maximum DM or $N_{e}$ value might have been obtained. No significant DM variation is evident in the non-eclipse phase for this pulsar.

\begin{figure*}[!htbp]
\begin{center}
\includegraphics[width=0.99\textwidth,angle=0]{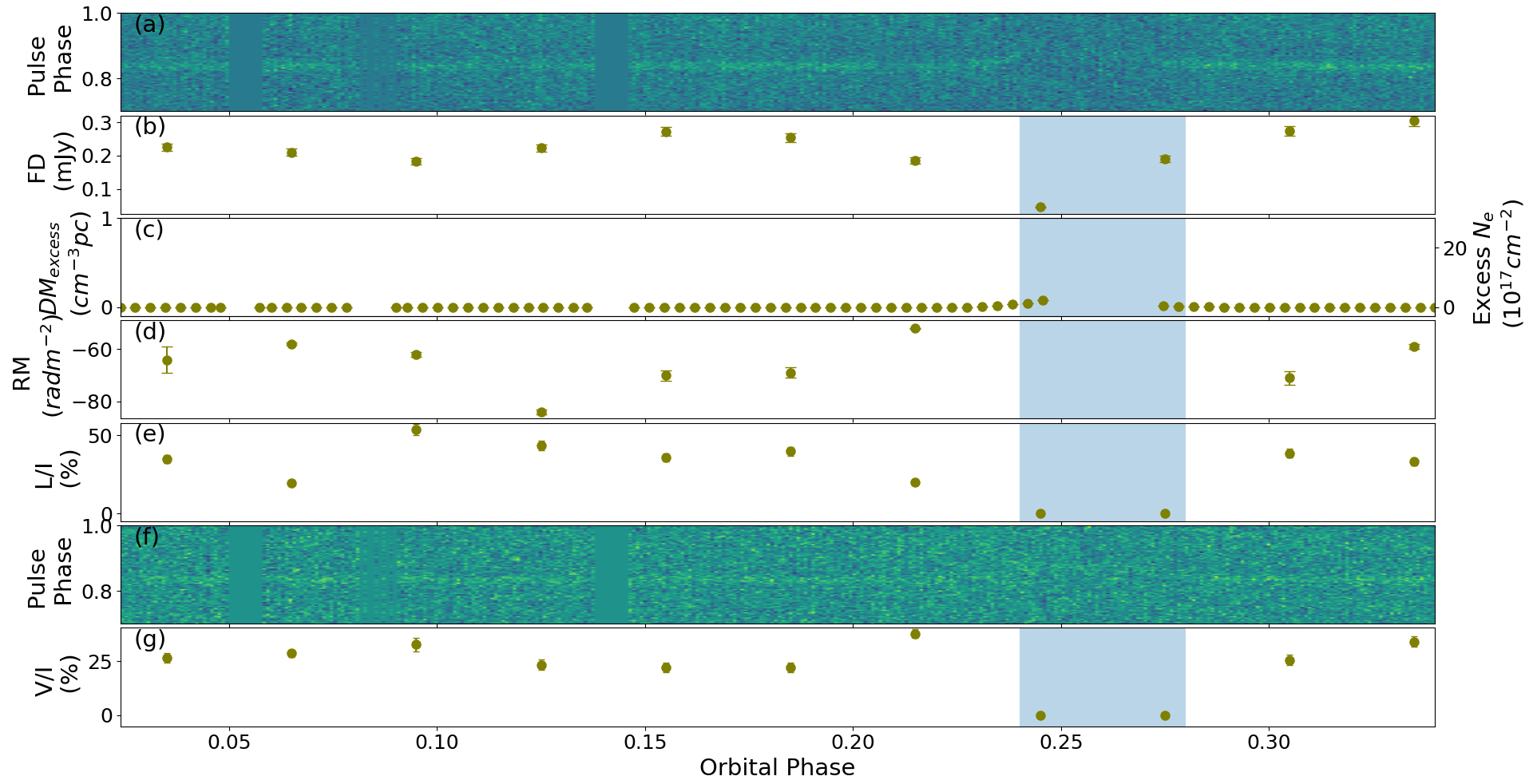}
\caption{Seven panels from top to bottom depict, (a) the total intensity, (b) flux density of total intensity, (c)  excess DM ($DM_{excess}$) along the line of sight, (d) RM, (e) linear polarization percentage, (f) circular polarization intensity, and (g) circular polarization percentage variation with orbital phase for PSR J1959$+$2048 on 27th January 2021, specifically for the component 2. The blue-shaded eclipse region is based on the depolarization width. The above quantities have been calculated for the 704-1900 MHz frequency range.}
\label{fig:J1959_plot}
\end{center}
\end{figure*}

For PSR J1959$+$2048 in all four epochs, we obtained only the lower limit of $N_{e}$ as a complete eclipse was observed throughout the band. Figure \ref{fig:J1959_plot}c showcases the DM variations and the corresponding orbital-dependent $N_{e}$ variation for PSR J1959$+$2048 for 27th January 2021. Table \ref{table2} lists the $N_{e}$ value obtained for these three MSPs.

\subsection{Variation of RM with orbital phase}
We investigated the RM change with orbital phase for the three pulsars in our sample, to determine the orbital phase dependent magnetic field. For PSR J0024$-$7204J, Figure \ref{fig:J0024_plot}d depicts the RM variation with the orbital phase. Notably, for the first eclipse covered, the RM value decreases until the eclipse ingress (exception being the point at orbital phase 0.65). In the eclipse region, no determination of RM is possible, as the pulsar undergoes depolarization, as seen for other BW MSPs \citep{J2051depol,J1748depol,J1748_nature}. On the other hand, we observed no reduction in the circular polarization intensity for PSR J0024$-$7204J in the eclipse region during the first eclipse. However, there is some decline in the circular polarization intensity during the second eclipse, as evident from Figure \ref{fig:J0024_plot}h. The non-eclipse phase RM is determined to be +20 $rad~m^{-2}$, spanning the orbital phase from 0.52 to 0.88.
With the MeerKAT telescope, an RM value of +24 $rad~m^{-2}$ is obtained after excluding the eclipse phase \citep{Meerkat_RM_J0024}. A study by \cite{parkes_RM_J0024}, using the central beam of the Parkes Multi-Beam Receiver with a central frequency of 1382 MHz and a bandwidth of 400 MHz, reported an RM value of $-$9 $rad~m^{-2}$ for this pulsar. Another study using the Parkes UWL receiver data yielded an RM value of +20 $rad~m^{-2}$ \citep{Parkes_UWL_RM_J0024}. However, it is not clear if, in the RM determination, the eclipse phase is excluded in these analyses. We utilized the data set from \cite{Parkes_UWL_RM_J0024} to investigate the orbital phase dependent RM variations, which was not explored by the authors. 

For PSR J1431$-$4715, Figure \ref{fig:J1431_plot}b shows the orbital phase-dependent RM variation, with RM values ranging from +23 $rad~m^{-2}$ to +33 $radm^{-2}$ during the non-eclipse phase. During the eclipse phase, we observed the depolarization of pulses. The RM value estimated in the non-eclipse phase (orbital phase spanning $\sim$0.62$-$0.94) on 23 November 2023 (as this epoch covers the non-eclipse phase for the maximum duration), is $+23 ~rad~m^{-2}$.
A previous polarization study of this pulsar was conducted using the MeerKAT radio telescope, revealing an RM value of +13.3 $rad~m^{-2}$ \citep{Meerkat_J1431_RM}. However, it is not specified whether this value was obtained by excluding the eclipse phase.

For PSR J1959$+$2048, Figure \ref{fig:J1959_plot}d shows the orbital phase-resolved RM variation for the component 2, for which we noted linear polarization. The RM values are measured outside the eclipse phase and range from $-$52 to +84 $rad~m^{-2}$. However, in the eclipse phase, specifically between 0.24$-$0.30 the linear polarization fraction is below the noise level, precluding RM determination. We also note that the circular polarization intensity is insignificant during the eclipse phase, between orbital phases 0.24$-$0.30. Using the complete non-eclipse phase data (orbital phase range $\sim$ 0.03-0.21), the RM value is estimated to be $-70 ~rad~m^{-2}$. A polarization study of this pulsar is previously carried out with the FAST radio telescope and reported an RM value of $-$70 $rad~m^{-2}$ \citep{J1959_fast_pol_21}. Nevertheless, it is still uncertain whether this measurement is conducted exclusively during the non-eclipse phase.

We also estimated the average parallel component of the ISM magnetic field using Equation \ref{magnetic_determination} (utilizing the non-eclipse phase RM and DM values) for the three MSPs in our sample. We obtained values of approximately 1 $\mu$G, 2.4 $\mu$G and  0.5 $\mu$G for PSR J0024$-$7204J, J1959$+$2048 and PSR J1431$-$4715, respectively.


\section{Discussion}
\label{discussion}

\subsection{Probing the eclipse mechanism from the cutoff frequency estimate for PSR J1431$-$4715}
\label{Probing the eclipse mechanism from the cutoff frequency estimate}
 We probed the eclipse mechanism for PSR J1431$-$4715, by applying the mechanisms proposed for frequency-dependent eclipsing in \cite{Thompson1992}, considering a cutoff frequency of 1251$\pm$80 MHz. 
The estimated lower limit of the $N_{e}$ is 9$\times10^{17} cm^{-2}$, as evident from Figure \ref{fig:J1431_plot}a. We calculated the eclipse radius ($R_{c}$) to be 1.35 $\times 10^{11}$ cm, by measuring the orbital phase range around the superior conjunction where excess electron density has been observed. The derived electron density $n_{e}$= $N_{e}/2R_{c}$ within the eclipse medium is 3.3 $\times 10^{6} cm^{-3}$.

Plasma frequency can cause an eclipse when the radio wave frequency is lower than the plasma frequency in the medium. Using the value of obtained $n_{e}$, we estimated the plasma frequency cutoff, $\mathrm{f_{p} = 8.5~ (\frac{n_{e}}{cm^{-3}})^{1/2} ~ kHz } = 15.44 ~\mathrm{MHz}$, to be significantly lower than the observed eclipse cutoff frequency.
Eclipse due to refraction can be ruled out, as the observed delays are in microseconds, whereas the required delays for refraction to be the main eclipse mechanism are in milliseconds \citep{Thompson1992}.
Scattering of radio waves causes pulse broadening, potentially leading to non-detection of pulsed emission if the broadening exceeds the spin period of pulsar, resulting in an eclipse. The effective pulse width, denoted as $W_{eff}$ is determined by $W_{eff} = \sqrt{W_{int}^{2} +W_{DM}^{2} +W_{scat}^{2} }$, where $W_{int}$ represents the intrinsic pulse width, $W_{DM}$
denotes the increase due to intra-channel dispersion smearing, and $W_{scat}$
accounts for broadening caused by scattering. 
For the estimation of scattering timescales, we have considered the DM of 59.55 $cm^{-3} ~pc$, which is the sum of non-eclipse phase DM = 59.35 $cm^{-3} ~pc$ and excess DM in the eclipse region, $DM_{excess}$  = 0.2 $cm^{-3} ~pc$. $W_{scat}$ is calculated to be 6.3 $\mu s$ at 1 GHz, using the expression derived by \cite{scattering_lewandowski} (Equation 7 of their paper), which is significantly less than the spin period of the pulsar. Consequently, scattering cannot account for the eclipse at 1251 MHz.
Eclipse due to free-free absorption can also be ruled out as solving for the optical depth ($\tau_{ff}$) equation \citep[Equation 11 of ][]{Thompson1992} with $\tau_{ff}>$  1, gives the condition T $\leq$ $747$ $f_{cl}^{2/3}$, where T is the temperature in the eclipse medium and $f_{cl}$ is the clumping factor. For the above condition to be satisfied either very low temperatures or a very high clumping factor is required, both of which are practically not feasible in the eclipse region as the pulsar radiation is itself intense enough to heat the plasma beyond this required temperature value.
Eclipse caused by induced Compton scattering can also be ruled out as the obtained optical depth for induced Compton scattering is less than 1 \citep[calculated using Equation 26 of][]{Thompson1992}.

Eclipse due to cyclotron absorption can also be ruled out. The optical depth for cyclotron absorption is given by Equation 43 of \cite{Thompson1992}. In that equation, the parameter m is the integer harmonic of the cyclotron frequency and is defined as m = $\nu_{c}/\nu_{b}$,  where $\nu_{b} =\frac{eB}{2\pi m_{e} c}$ is the cyclotron frequency (e is the charge of electron, B is the magnetic field strength and c is the speed of light) and $\nu_{c}$ is the cutoff frequency which is estimated to be 1251 MHz for PSR J1431$-$4715. The value of m depends on the magnetic field strength and its variation. We assumed the magnetic field to be equal to the characteristic magnetic field ($B_{E}\sim$ 10 $\mathrm{Gauss}$), calculated using the pressure balance between pulsar wind energy density ($U_{E}=\frac{\dot{E}}{4\pi ca^{2}}$) and the stellar wind energy density of the companion ($\frac{B_{E}^{2}}{8\pi}$), where a is the distance between the pulsar and the companion ($a \sim 3.0 R _{\odot}$, assuming the inclination angle of $60^\circ$). Using this magnetic field value and the $\nu_{c}$, the estimated value of m is 44. For simplicity, we consider the scale length of electron density variations and magnetic field variations as similar. Thus, the quantity $n_e \times L_{B}$, \citep[in Equation 43 of][]{Thompson1992} is taken to be equal to the average electron column density ($N_{e}$). Using these parameters \citep[in Equation 43 of][and taking $\alpha = \pi/2$, since considering $\alpha$ less than this value results in a higher value of T]{Thompson1992}, we find that for cyclotron absorption to be significant, the temperature above $10^{7}$ K in the eclipse medium is required. However, cyclotron absorption is valid in the non-relativistic regime for T $< \frac{m_{e}c^{2}}{2km^{3}}$ = 3.4 $\times 10^{4}$ K \citep[Equation C2 of][]{Thompson1992}. Thus, cyclotron absorption can be ruled out as the cause of the eclipse for PSR J1431$-$4715.

Therefore, we conclude that synchrotron absorption is the major eclipse mechanism for PSR J1431$-$4715. For PSR J0024$-$7204J, constraints on the eclipse cutoff frequency prevent thorough exploration of the eclipse mechanism. Meanwhile, at 1.5 GHz for PSR J1959+2048, \cite{Thompson1992} extensively studied the eclipse mechanism, inferring pulse smearing as the most plausible mechanism at this frequency.

\subsection{Depolarization of pulsed emission in the eclipse phase}
\label{Depolarisation of pulsar in the eclipse phase}

Understanding the magnetic field within the eclipse region is crucial for confirming the cyclotron-synchrotron absorption mechanism in BW MSPs. Efforts have been made in the past to determine the magnetic field in the eclipse region for spider MSPs. However, until now, it has only been possible to ascertain the value of the magnetic field at the eclipse boundary, not at its centre. 
For example, in a recent study by \cite{J2051depol}, the magnetic field for J2051$-$0827 was estimated to be 0.1 G at the eclipse boundary. Meanwhile, using the plasma lensing technique, \cite{Li2019} estimated the magnetic field for PSR J1959$+$2048 to be less than 0.02 G at the eclipse boundary. 
The expected magnetic field \citep[characteristic magnetic field estimated by pressure balance between the pulsar energy density and the stellar wind energy density of the companion, using Equation 35 of][]{Thompson1992} at the eclipse centre is approximately 10 G. The difference between the observed and expected values might arise because the magnetic field measured by \cite{Li2019} and \cite{J2051depol} is not at the eclipse centre, but is measured at the eclipse boundary.
The magnetic field at the eclipse center remains uncertain either because the cutoff frequency for pulsar exceeds the observing band or during the eclipse phase reduction of linear polarization is observed (also known as depolarization). 
For example, depolarization in the eclipse medium is observed for PSR J2051$-$0827 by \cite{J2051depol}, for PSR J2256$-$1024 by \cite{J2256depol}, and for J1748$-$2446A by \cite{J1748depol,J1748_nature}. The observed depolarization could be caused by small-scale spatial fluctuations in electron density or magnetic field within the eclipse medium, resulting in rapid temporal variations in RM and averaging over a few sub-integrations can lead to a reduction in the linear polarization intensity \citep{J1748depol}.

The depolarization is observed for $\sim$ 50 $\%$ of the orbit at 921 MHz (with a bandwidth of 374 MHz) for PSR J0024$-$7204J (approximately from 1.0$\pm0.02$ to $1.5\pm0.02$ orbital phase, as evident from Figure \ref{fig:J0024_plot}f). For PSR J1431$-$4715 the depolarization at 1.4 GHz (with a bandwidth of 350 MHz) is observed for about 22 $\%$ of the orbit (approximately $0.15\pm0.02$ to $0.37\pm0.01$ orbital phase, as evident from Figure \ref{fig:J1431_plot}d). For PSR J1959$+$2048, depolarization is observed for 6$\%$ of the orbit (from orbital phase 0.24$\pm$0.02 to 0.30$\pm$0.02) at 1.3 GHz (with a bandwidth of 1200 MHz). Although a complete eclipse is noted for this MSP till the frequency up to which the pulsar is detected in the non-eclipse phase, we observed that between orbital phases 0.255 to 0.285, where total intensity is detected, the linear polarization fraction is zero. Therefore, we can conclude that pulse depolarization during the eclipse phase is evident for this MSP as well.  In the previous studies, \cite{J2051depol} and \cite{J1748_nature} observed depolarization for $\sim$ 10 $\%$ and $\sim$ 50 $\%$ of the orbit, at 1250 MHz and 1.5 GHz for PSR J2051$-$0827 and PSR J1748$-$2446A, respectively. 

We also observed frequency-dependent depolarization for PSR J1431$-$4715 and found it to be more pronounced at lower frequencies. Depolarization is observed for approximately 33$\%$ of the orbit in the 704$-$1251 MHz range and 22$\%$ of the orbit in the 1251$-$1600 MHz range. Frequency-dependent depolarization is also observed by \cite{J1748depol} in PSR J1748–2446A. 

Considering the random RM variations following a normal distribution with a standard deviation of $\sigma_{RM}$, the ratio of the observed linear polarization ($L_{obs}$) to the actual linear polarization (L) is given as, $L_{obs}/L = e^{-2\sigma_{RM}^{2}\lambda^{4}}$ \citep{J1748_nature}. For depolarization, $\sigma_{RM}$ values greater than 9 $rad~m^{-2}$ at 921 MHz, 25 $rad~m^{-2}$ at 1.4 GHz and 18 $rad~m^{-2}$ at 1.3 GHz are required. 
From Figures \ref{fig:J1431_plot}, and \ref{fig:J1959_plot}, it can be noted that the depolarization width is similar to the eclipse width (estimated by the orbital phase range around the superior conjunction where excess electron density has been observed).

We were unable to estimate the magnetic field strength at the eclipse boundary, as conducted by \cite{J2051depol} for PSR J2051$-$0827, due to the lack of sensitivity in our observations to estimate the RM values at the eclipse boundary.

\section{Summary}
\label{summary}
This paper presents the cutoff frequency and orbital phase-dependent polarization study for three BW MSPs, namely PSR J0024$-$7204J, PSR J1731$-$4715 and PSR J1959$+$2048. We concluded that the eclipse cutoff frequency for PSR J0024$-$7204J, PSR J1731$-$4715 could vary with time, along with the temporal variation of $N_{e}$ in the eclipse region. Such variations could indicate that the eclipse environment is dynamically evolving, as observed for PSR J1544$+$4937 by \cite{kumari2024}. Using the cutoff frequency estimate of 1251 MHz for PSR J1731$-$4715, we concluded that synchrotron absorption by trans-relativistic free electrons in the eclipse medium is the major eclipse mechanism. Additionally, we observed an unusual increase in total intensity in the eclipse region for PSR J0024$-$7204J, which is generally not seen for other BW MSPs. In other BW MSPs, a reduction of total intensity in the eclipse region is observed in previous studies. In the orbital phase-dependent polarization study, we observed the depolarization of pulses in the eclipse phase for PSR J0024$-$7204J, J1731$-$4715 and PSR J1959$+$2048. The cause of this depolarization could be rapid fluctuations of RM in the eclipse medium. In the past, orbital phase resolved studies have been conducted for only 3 BW MSPs, and with this study we have doubled the sample size. Conducting more such studies of these systems would further increase the understanding of the eclipse environment and would allow to probe the underlying physics responsible for it.


\begin{acknowledgments}
We acknowledge the support of the Department of Atomic Energy, Government of India, under project no.12-R\&D-TFR-5.02-0700. Murriyang, the Parkes radio telescope, is part of the Australia Telescope National Facility (https://ror.org/05qajvd42) which is funded by the Australian Government for operation as a National Facility managed by CSIRO.
\end{acknowledgments}

\bibliography{citation.bib}{}

\begin{thebibliography}{}
\expandafter\ifx\csname natexlab\endcsname\relax\def\natexlab#1{#1}\fi
\providecommand{\url}[1]{\href{#1}{#1}}
\providecommand{\dodoi}[1]{doi:~\href{http://doi.org/#1}{\nolinkurl{#1}}}
\providecommand{\doeprint}[1]{\href{http://ascl.net/#1}{\nolinkurl{http://ascl.net/#1}}}
\providecommand{\doarXiv}[1]{\href{https://arxiv.org/abs/#1}{\nolinkurl{https://arxiv.org/abs/#1}}}

\bibitem[{{Abbate} {et~al.}(2020){Abbate}, {Possenti}, {Tiburzi}, {Barr}, {van Straten}, {Ridolfi}, \& {Freire}}]{parkes_RM_J0024}
{Abbate}, F., {Possenti}, A., {Tiburzi}, C., {et~al.} 2020, Nature Astronomy, 4, 704, \dodoi{10.1038/s41550-020-1030-6}

\bibitem[{{Abbate} {et~al.}(2023){Abbate}, {Possenti}, {Ridolfi}, {Venkatraman Krishnan}, {Buchner}, {Barr}, {Bailes}, {Kramer}, {Cameron}, {Parthasarathy}, {van Straten}, {Chen}, {Camilo}, {Padmanabh}, {Mao}, {Freire}, {Ransom}, {Vleeschower}, {Geyer}, \& {Zhang}}]{Meerkat_RM_J0024}
{Abbate}, F., {Possenti}, A., {Ridolfi}, A., {et~al.} 2023, \mnras, 518, 1642, \dodoi{10.1093/mnras/stac3248}

\bibitem[{{Archibald} {et~al.}(2009){Archibald}, {Stairs}, {Ransom}, {Kaspi}, {Kondratiev}, {Lorimer}, {McLaughlin}, {Boyles}, {Hessels}, {Lynch}, {van Leeuwen}, {Roberts}, {Jenet}, {Champion}, {Rosen}, {Barlow}, {Dunlap}, \& {Remillard}}]{Roche_lobe_eclipse}
{Archibald}, A.~M., {Stairs}, I.~H., {Ransom}, S.~M., {et~al.} 2009, Science, 324, 1411, \dodoi{10.1126/science.1172740}

\bibitem[{{Bilous} {et~al.}(2019){Bilous}, {Ransom}, \& {Demorest}}]{B1744_plasma_lensing}
{Bilous}, A.~V., {Ransom}, S.~M., \& {Demorest}, P. 2019, \apj, 877, 125, \dodoi{10.3847/1538-4357/ab16dd}

\bibitem[{{Crowter} {et~al.}(2020){Crowter}, {Stairs}, {McPhee}, {Archibald}, {Boyles}, {Hessels}, {Karako-Argaman}, {Lorimer}, {Lynch}, {McLaughlin}, {Ransom}, {Roberts}, {Stovall}, \& {van Leeuwen}}]{J2256depol}
{Crowter}, K., {Stairs}, I.~H., {McPhee}, C.~A., {et~al.} 2020, \mnras, 495, 3052, \dodoi{10.1093/mnras/staa933}

\bibitem[{{Eggleton}(1983)}]{Roche_eggleton}
{Eggleton}, P.~P. 1983, \apj, 268, 368, \dodoi{10.1086/160960}

\bibitem[{{Freire}(2005)}]{J0024_previous_study}
{Freire}, P.~C.~C. 2005, in Astronomical Society of the Pacific Conference Series, Vol. 328, Binary Radio Pulsars, ed. F.~A. {Rasio} \& I.~H. {Stairs}, 405, \dodoi{10.48550/arXiv.astro-ph/0404105}

\bibitem[{{Fruchter} {et~al.}(1988){Fruchter}, {Stinebring}, \& {Taylor}}]{Fruchter1988a}
{Fruchter}, A.~S., {Stinebring}, D.~R., \& {Taylor}, J.~H. 1988, \nat, 333, 237, \dodoi{10.1038/333237a0}

\bibitem[{{Guillemot} {et~al.}(2019){Guillemot}, {Octau}, {Cognard}, {Desvignes}, {Freire}, {Smith}, {Theureau}, \& {Burnett}}]{RochelobeJ2055}
{Guillemot}, L., {Octau}, F., {Cognard}, I., {et~al.} 2019, \aap, 629, A92, \dodoi{10.1051/0004-6361/201936015}

\bibitem[{Hobbs {et~al.}(2006)Hobbs, Edwards, \& Manchester}]{hobbs2006tempo2}
Hobbs, G., Edwards, R., \& Manchester, R. 2006, Monthly Notices of the Royal Astronomical Society, 369, 655

\bibitem[{{Hobbs} {et~al.}(2020){Hobbs}, {Manchester}, {Dunning}, {Jameson}, {Roberts}, {George}, {Green}, {Tuthill}, {Toomey}, {Kaczmarek}, {Mader}, {Marquarding}, {Ahmed}, {Amy}, {Bailes}, {Beresford}, {Bhat}, {Bock}, {Bourne}, {Bowen}, {Brothers}, {Cameron}, {Carretti}, {Carter}, {Castillo}, {Chekkala}, {Cheng}, {Chung}, {Craig}, {Dai}, {Dawson}, {Dempsey}, {Doherty}, {Dong}, {Edwards}, {Ergesh}, {Gao}, {Han}, {Hayman}, {Indermuehle}, {Jeganathan}, {Johnston}, {Kanoniuk}, {Kesteven}, {Kramer}, {Leach}, {Mcintyre}, {Moss}, {Os{\l}owski}, {Phillips}, {Pope}, {Preisig}, {Price}, {Reeves}, {Reilly}, {Reynolds}, {Robishaw}, {Roush}, {Ruckley}, {Sadler}, {Sarkissian}, {Severs}, {Shannon}, {Smart}, {Smith}, {Smith}, {Sobey}, {Staveley-Smith}, {Tzioumis}, {van Straten}, {Wang}, {Wen}, \& {Whiting}}]{UWL_receiver}
{Hobbs}, G., {Manchester}, R.~N., {Dunning}, A., {et~al.} 2020, \pasa, 37, e012, \dodoi{10.1017/pasa.2020.2}

\bibitem[{Kansabanik {et~al.}(2021)Kansabanik, Bhattacharyya, Roy, \& Stappers}]{kansabanik2021unraveling}
Kansabanik, D., Bhattacharyya, B., Roy, J., \& Stappers, B. 2021, The Astrophysical Journal, 920, 58

\bibitem[{{Khechinashvili} {et~al.}(2000){Khechinashvili}, {Melikidze}, \& {Gil}}]{J1959_previous_study}
{Khechinashvili}, D.~G., {Melikidze}, G.~I., \& {Gil}, J.~A. 2000, \apj, 541, 335, \dodoi{10.1086/309408}

\bibitem[{{Kijak} {et~al.}(2011){Kijak}, {Lewandowski}, {Maron}, {Gupta}, \& {Jessner}}]{high_frequency_turnover_b}
{Kijak}, J., {Lewandowski}, W., {Maron}, O., {Gupta}, Y., \& {Jessner}, A. 2011, \aap, 531, A16, \dodoi{10.1051/0004-6361/201014274}

\bibitem[{Kramer(2005)}]{LorimerKramer}
Kramer, L.~. 2005, Handbook of Pulsar Astronomy (Cambridge University Press)

\bibitem[{{Kudale} {et~al.}(2020){Kudale}, {Roy}, {Bhattacharyya}, {Stappers}, \& {Chengalur}}]{Kudale2020}
{Kudale}, S., {Roy}, J., {Bhattacharyya}, B., {Stappers}, B., \& {Chengalur}, J. 2020, \apj, 900, 194, \dodoi{10.3847/1538-4357/aba902}

\bibitem[{{Kumari} {et~al.}(2024){Kumari}, {Bhattacharyya}, {Sharan}, {Kansabanik}, {Stappers}, \& {Roy}}]{kumari2024}
{Kumari}, S., {Bhattacharyya}, B., {Sharan}, R., {et~al.} 2024, \apj, 961, 155, \dodoi{10.3847/1538-4357/ad0b83}

\bibitem[{{Lewandowski} {et~al.}(2015){Lewandowski}, {Kowali{\'n}ska}, \& {Kijak}}]{scattering_lewandowski}
{Lewandowski}, W., {Kowali{\'n}ska}, M., \& {Kijak}, J. 2015, \mnras, 449, 1570, \dodoi{10.1093/mnras/stv385}

\bibitem[{{Li} {et~al.}(2023){Li}, {Bilous}, {Ransom}, {Main}, \& {Yang}}]{J1748_nature}
{Li}, D., {Bilous}, A., {Ransom}, S., {Main}, R., \& {Yang}, Y.-P. 2023, \nat, 618, 484, \dodoi{10.1038/s41586-023-05983-z}

\bibitem[{{Li} {et~al.}(2019){Li}, {Lin}, {Main}, {Pen}, {van Kerkwijk}, \& {Yang}}]{Li2019}
{Li}, D., {Lin}, F.~X., {Main}, R., {et~al.} 2019, \mnras, 484, 5723, \dodoi{10.1093/mnras/stz374}

\bibitem[{{Lin} {et~al.}(2021){Lin}, {Main}, {Verbiest}, {Kramer}, \& {Shaifullah}}]{J2051_plasma_lensing}
{Lin}, F.~X., {Main}, R.~A., {Verbiest}, J.~P.~W., {Kramer}, M., \& {Shaifullah}, G. 2021, \mnras, 506, 2824, \dodoi{10.1093/mnras/stab1811}

\bibitem[{{Main} {et~al.}(2018){Main}, {Yang}, {Chan}, {Li}, {Lin}, {Mahajan}, {Pen}, {Vanderlinde}, \& {van Kerkwijk}}]{J1959_plasma_lensing}
{Main}, R., {Yang}, I.~S., {Chan}, V., {et~al.} 2018, \nat, 557, 522, \dodoi{10.1038/s41586-018-0133-z}

\bibitem[{{Miraval Zanon} {et~al.}(2018){Miraval Zanon}, {Burgay}, {Possenti}, \& {Ridolfi}}]{J1431_previous_study}
{Miraval Zanon}, A., {Burgay}, M., {Possenti}, A., \& {Ridolfi}, A. 2018, in Journal of Physics Conference Series, Vol. 956, Journal of Physics Conference Series, 012004, \dodoi{10.1088/1742-6596/956/1/012004}

\bibitem[{{Nan} {et~al.}(2011){Nan}, {Li}, {Jin}, {Wang}, {Zhu}, {Zhu}, {Zhang}, {Yue}, \& {Qian}}]{FAST_telescope}
{Nan}, R., {Li}, D., {Jin}, C., {et~al.} 2011, International Journal of Modern Physics D, 20, 989, \dodoi{10.1142/S0218271811019335}

\bibitem[{{Polzin} {et~al.}(2019){Polzin}, {Breton}, {Stappers}, {Bhattacharyya}, {Janssen}, {Os{\l}owski}, {Roberts}, \& {Sobey}}]{PolzinJ2051}
{Polzin}, E.~J., {Breton}, R.~P., {Stappers}, B.~W., {et~al.} 2019, \mnras, 490, 889, \dodoi{10.1093/mnras/stz2579}

\bibitem[{{Polzin} {et~al.}(2018){Polzin}, {Breton}, {Clarke}, {Kondratiev}, {Stappers}, {Hessels}, {Bassa}, {Broderick}, {Grie{\ss}meier}, {Sobey}, {ter Veen}, {van Leeuwen}, \& {Weltevrede}}]{PolzinJ1810}
{Polzin}, E.~J., {Breton}, R.~P., {Clarke}, A.~O., {et~al.} 2018, \mnras, 476, 1968, \dodoi{10.1093/mnras/sty349}

\bibitem[{{Rajwade} {et~al.}(2016){Rajwade}, {Lorimer}, \& {Anderson}}]{high_frequency_turnover}
{Rajwade}, K., {Lorimer}, D.~R., \& {Anderson}, L.~D. 2016, \mnras, 455, 493, \dodoi{10.1093/mnras/stv2334}

\bibitem[{Roberts(2012)}]{roberts2012surrounded}
Roberts, M.~S. 2012, Proceedings of the International Astronomical Union, 8, 127

\bibitem[{{Spiewak} {et~al.}(2022){Spiewak}, {Bailes}, {Miles}, {Parthasarathy}, {Reardon}, {Shamohammadi}, {Shannon}, {Bhat}, {Buchner}, {Cameron}, {Camilo}, {Geyer}, {Johnston}, {Karastergiou}, {Keith}, {Kramer}, {Serylak}, {van Straten}, {Theureau}, \& {Venkatraman Krishnan}}]{Meerkat_J1431_RM}
{Spiewak}, R., {Bailes}, M., {Miles}, M.~T., {et~al.} 2022, \pasa, 39, e027, \dodoi{10.1017/pasa.2022.19}

\bibitem[{{Thompson} {et~al.}(1994){Thompson}, {Blandford}, {Evans}, \& {Phinney}}]{Thompson1992}
{Thompson}, C., {Blandford}, R.~D., {Evans}, C.~R., \& {Phinney}, E.~S. 1994, \apj, 422, 304, \dodoi{10.1086/173728}

\bibitem[{{van Straten}(2013)}]{pcm_file}
{van Straten}, W. 2013, \apjs, 204, 13, \dodoi{10.1088/0067-0049/204/1/13}

\bibitem[{{Wang} {et~al.}(2023{\natexlab{a}}){Wang}, {Han}, {Xu}, {Wang}, {Yan}, {Jing}, {Su}, {Zhou}, \& {Wang}}]{J1959_fast_pol_21}
{Wang}, P.~F., {Han}, J.~L., {Xu}, J., {et~al.} 2023{\natexlab{a}}, Research in Astronomy and Astrophysics, 23, 104002, \dodoi{10.1088/1674-4527/acea1f}

\bibitem[{{Wang} {et~al.}(2023{\natexlab{b}}){Wang}, {Wang}, {Li}, {Yao}, {Manchester}, {Hobbs}, {Wang}, {Dai}, {Xu}, {Luo}, {Feng}, {Wang}, {Li}, {Yu}, {Du}, {Niu}, {Zhang}, \& {Zhang}}]{J2051depol}
{Wang}, S.~Q., {Wang}, J.~B., {Li}, D.~Z., {et~al.} 2023{\natexlab{b}}, \apj, 955, 36, \dodoi{10.3847/1538-4357/acea81}

\bibitem[{{You} {et~al.}(2018){You}, {Manchester}, {Coles}, {Hobbs}, \& {Shannon}}]{J1748depol}
{You}, X.~P., {Manchester}, R.~N., {Coles}, W.~A., {Hobbs}, G.~B., \& {Shannon}, R. 2018, \apj, 867, 22, \dodoi{10.3847/1538-4357/aadee0}

\bibitem[{{Zhang} {et~al.}(2019){Zhang}, {Hobbs}, {Manchester}, {Li}, {Wang}, {Dai}, {Wang}, {Kaczmarek}, {Cameron}, {Toomey}, {Zhu}, {Zhi}, {Miao}, {Yuan}, {Zhang}, \& {Tao}}]{Parkes_UWL_RM_J0024}
{Zhang}, L., {Hobbs}, G., {Manchester}, R.~N., {et~al.} 2019, \apjl, 885, L37, \dodoi{10.3847/2041-8213/ab5218}

\end{thebibliography}
\bibliographystyle{aasjournal}

\end{document}